\hsize=31pc
\vsize=49pc
\lineskip=0pt
\parskip=0pt plus 1pt
\hfuzz=1pt
\vfuzz=2pt
\pretolerance=2500
\tolerance=5000
\vbadness=5000
\hbadness=5000
\widowpenalty=500
\clubpenalty=200
\brokenpenalty=500
\predisplaypenalty=200
\voffset=-1pc
\nopagenumbers
\catcode`@=11
\newif\ifams
\amsfalse 
%
%
%
\newfam\bdifam
\newfam\bsyfam
\newfam\bssfam
\newfam\msafam
\newfam\msbfam
\newif\ifxxpt
\newif\ifxviipt
\newif\ifxivpt
\newif\ifxiipt
\newif\ifxipt
\newif\ifxpt
\newif\ifixpt
\newif\ifviiipt
\newif\ifviipt
\newif\ifvipt
\newif\ifvpt
%
%
\def\headsize#1#2{\def\headb@seline{#2}%
                \ifnum#1=20\def\HEAD{twenty}%
                           \def\smHEAD{twelve}%
                           \def\vsHEAD{nine}%
                           \ifxxpt\else\xdef\f@ntsize{\HEAD}%
                           \def\m@g{4}\def\s@ze{20.74}%
                           \loadheadfonts\xxpttrue\fi
                           \ifxiipt\else\xdef\f@ntsize{\smHEAD}%
                           \def\m@g{1}\def\s@ze{12}%
                           \loadxiiptfonts\xiipttrue\fi
                           \ifixpt\else\xdef\f@ntsize{\vsHEAD}%
                           \def\s@ze{9}%
                           \loadsmallfonts\ixpttrue\fi
                      \else
                \ifnum#1=17\def\HEAD{seventeen}%
                           \def\smHEAD{eleven}%
                           \def\vsHEAD{eight}%
                           \ifxviipt\else\xdef\f@ntsize{\HEAD}%
                           \def\m@g{3}\def\s@ze{17.28}%
                           \loadheadfonts\xviipttrue\fi
                           \ifxipt\else\xdef\f@ntsize{\smHEAD}%
                           \loadxiptfonts\xipttrue\fi
                           \ifviiipt\else\xdef\f@ntsize{\vsHEAD}%
                           \def\s@ze{8}%
                           \loadsmallfonts\viiipttrue\fi
                      \else\def\HEAD{fourteen}%
                           \def\smHEAD{ten}%
                           \def\vsHEAD{seven}%
                           \ifxivpt\else\xdef\f@ntsize{\HEAD}%
                           \def\m@g{2}\def\s@ze{14.4}%
                           \loadheadfonts\xivpttrue\fi
                           \ifxpt\else\xdef\f@ntsize{\smHEAD}%
                           \def\s@ze{10}%
                           \loadxptfonts\xpttrue\fi
                           \ifviipt\else\xdef\f@ntsize{\vsHEAD}%
                           \def\s@ze{7}%
                           \loadviiptfonts\viipttrue\fi
                \ifnum#1=14\else
                \message{Header size should be 20, 17 or 14 point
                              will now default to 14pt}\fi
                \fi\fi\headfonts}
%
%
\def\textsize#1#2{\def\textb@seline{#2}%
                 \ifnum#1=12\def\TEXT{twelve}%
                           \def\smTEXT{eight}%
                           \def\vsTEXT{six}%
                           \ifxiipt\else\xdef\f@ntsize{\TEXT}%
                           \def\m@g{1}\def\s@ze{12}%
                           \loadxiiptfonts\xiipttrue\fi
                           \ifviiipt\else\xdef\f@ntsize{\smTEXT}%
                           \def\s@ze{8}%
                           \loadsmallfonts\viiipttrue\fi
                           \ifvipt\else\xdef\f@ntsize{\vsTEXT}%
                           \def\s@ze{6}%
                           \loadviptfonts\vipttrue\fi
                      \else
                \ifnum#1=11\def\TEXT{eleven}%
                           \def\smTEXT{seven}%
                           \def\vsTEXT{five}%
                           \ifxipt\else\xdef\f@ntsize{\TEXT}%
                           \def\s@ze{11}%
                           \loadxiptfonts\xipttrue\fi
                           \ifviipt\else\xdef\f@ntsize{\smTEXT}%
                           \loadviiptfonts\viipttrue\fi
                           \ifvpt\else\xdef\f@ntsize{\vsTEXT}%
                           \def\s@ze{5}%
                           \loadvptfonts\vpttrue\fi
                      \else\def\TEXT{ten}%
                           \def\smTEXT{seven}%
                           \def\vsTEXT{five}%
                           \ifxpt\else\xdef\f@ntsize{\TEXT}%
                           \loadxptfonts\xpttrue\fi
                           \ifviipt\else\xdef\f@ntsize{\smTEXT}%
                           \def\s@ze{7}%
                           \loadviiptfonts\viipttrue\fi
                           \ifvpt\else\xdef\f@ntsize{\vsTEXT}%
                           \def\s@ze{5}%
                           \loadvptfonts\vpttrue\fi
                \ifnum#1=10\else
                \message{Text size should be 12, 11 or 10 point
                              will now default to 10pt}\fi
                \fi\fi\textfonts}
%
%
\def\smallsize#1#2{\def\smallb@seline{#2}%
                 \ifnum#1=10\def\SMALL{ten}%
                           \def\smSMALL{seven}%
                           \def\vsSMALL{five}%
                           \ifxpt\else\xdef\f@ntsize{\SMALL}%
                           \loadxptfonts\xpttrue\fi
                           \ifviipt\else\xdef\f@ntsize{\smSMALL}%
                           \def\s@ze{7}%
                           \loadviiptfonts\viipttrue\fi
                           \ifvpt\else\xdef\f@ntsize{\vsSMALL}%
                           \def\s@ze{5}%
                           \loadvptfonts\vpttrue\fi
                       \else
                 \ifnum#1=9\def\SMALL{nine}%
                           \def\smSMALL{six}%
                           \def\vsSMALL{five}%
                           \ifixpt\else\xdef\f@ntsize{\SMALL}%
                           \def\s@ze{9}%
                           \loadsmallfonts\ixpttrue\fi
                           \ifvipt\else\xdef\f@ntsize{\smSMALL}%
                           \def\s@ze{6}%
                           \loadviptfonts\vipttrue\fi
                           \ifvpt\else\xdef\f@ntsize{\vsSMALL}%
                           \def\s@ze{5}%
                           \loadvptfonts\vpttrue\fi
                       \else
                           \def\SMALL{eight}%
                           \def\smSMALL{six}%
                           \def\vsSMALL{five}%
                           \ifviiipt\else\xdef\f@ntsize{\SMALL}%
                           \def\s@ze{8}%
                           \loadsmallfonts\viiipttrue\fi
                           \ifvipt\else\xdef\f@ntsize{\smSMALL}%
                           \def\s@ze{6}%
                           \loadviptfonts\vipttrue\fi
                           \ifvpt\else\xdef\f@ntsize{\vsSMALL}%
                           \def\s@ze{5}%
                           \loadvptfonts\vpttrue\fi
                 \ifnum#1=8\else\message{Small size should be 10, 9 or
                            8 point will now default to 8pt}\fi
                \fi\fi\smallfonts}
\def\F@nt{\expandafter\font\csname}
\def\Sk@w{\expandafter\skewchar\csname}
\def\@nd{\endcsname}
\def\@step#1{ scaled \magstep#1}
\def\@half{ scaled \magstephalf}
\def\@t#1{ at #1pt}
%
%
\def\loadheadfonts{\bigf@nts
\F@nt \f@ntsize bdi\@nd=cmmib10 \@t{\s@ze}%
\Sk@w \f@ntsize bdi\@nd='177
\F@nt \f@ntsize bsy\@nd=cmbsy10 \@t{\s@ze}%
\Sk@w \f@ntsize bsy\@nd='60
\F@nt \f@ntsize bss\@nd=cmssbx10 \@t{\s@ze}}
%
%
\def\loadxiiptfonts{\bigf@nts
\F@nt \f@ntsize bdi\@nd=cmmib10 \@step{\m@g}%
\Sk@w \f@ntsize bdi\@nd='177
\F@nt \f@ntsize bsy\@nd=cmbsy10 \@step{\m@g}%
\Sk@w \f@ntsize bsy\@nd='60
\F@nt \f@ntsize bss\@nd=cmssbx10 \@step{\m@g}}
%
%
\def\loadxiptfonts{%
\font\elevenrm=cmr10 \@half
\font\eleveni=cmmi10 \@half
\skewchar\eleveni='177
\font\elevensy=cmsy10 \@half
\skewchar\elevensy='60
\font\elevenex=cmex10 \@half
\font\elevenit=cmti10 \@half
\font\elevensl=cmsl10 \@half
\font\elevenbf=cmbx10 \@half
\font\eleventt=cmtt10 \@half
\ifams\font\elevenmsa=msam10 \@half
\font\elevenmsb=msbm10 \@half\else\fi
\font\elevenbdi=cmmib10 \@half
\skewchar\elevenbdi='177
\font\elevenbsy=cmbsy10 \@half
\skewchar\elevenbsy='60
\font\elevenbss=cmssbx10 \@half}
%
%
\def\loadxptfonts{%
\font\tenbdi=cmmib10
\skewchar\tenbdi='177
\font\tenbsy=cmbsy10
\skewchar\tenbsy='60
\ifams\font\tenmsa=msam10
\font\tenmsb=msbm10\else\fi
\font\tenbss=cmssbx10}%
%
%
\def\loadsmallfonts{\smallf@nts
\ifams
\F@nt \f@ntsize ex\@nd=cmex\s@ze
\else
\F@nt \f@ntsize ex\@nd=cmex10\fi
\F@nt \f@ntsize it\@nd=cmti\s@ze
\F@nt \f@ntsize sl\@nd=cmsl\s@ze
\F@nt \f@ntsize tt\@nd=cmtt\s@ze}
%
%
\def\loadviiptfonts{%
\font\sevenit=cmti7
\font\sevensl=cmsl8 at 7pt
\ifams\font\sevenmsa=msam7
\font\sevenmsb=msbm7
\font\sevenex=cmex7
\font\sevenbsy=cmbsy7
\font\sevenbdi=cmmib7\else
\font\sevenex=cmex10
\font\sevenbsy=cmbsy10 at 7pt
\font\sevenbdi=cmmib10 at 7pt\fi
\skewchar\sevenbsy='60
\skewchar\sevenbdi='177
\font\sevenbss=cmssbx10 at 7pt}%
%
%
\def\loadviptfonts{\smallf@nts
\ifams\font\sixex=cmex7 at 6pt\else
\font\sixex=cmex10\fi
\font\sixit=cmti7 at 6pt}
%
%
\def\loadvptfonts{%
\font\fiveit=cmti7 at 5pt
\ifams\font\fiveex=cmex7 at 5pt
\font\fivebdi=cmmib5
\font\fivebsy=cmbsy5
\font\fivemsa=msam5
\font\fivemsb=msbm5\else
\font\fiveex=cmex10
\font\fivebdi=cmmib10 at 5pt
\font\fivebsy=cmbsy10 at 5pt\fi
\skewchar\fivebdi='177
\skewchar\fivebsy='60
\font\fivebss=cmssbx10 at 5pt}
\def\bigf@nts{%
\F@nt \f@ntsize rm\@nd=cmr10 \@step{\m@g}%
\F@nt \f@ntsize i\@nd=cmmi10 \@step{\m@g}%
\Sk@w \f@ntsize i\@nd='177
\F@nt \f@ntsize sy\@nd=cmsy10 \@step{\m@g}%
\Sk@w \f@ntsize sy\@nd='60
\F@nt \f@ntsize ex\@nd=cmex10 \@step{\m@g}%
\F@nt \f@ntsize it\@nd=cmti10 \@step{\m@g}%
\F@nt \f@ntsize sl\@nd=cmsl10 \@step{\m@g}%
\F@nt \f@ntsize bf\@nd=cmbx10 \@step{\m@g}%
\F@nt \f@ntsize tt\@nd=cmtt10 \@step{\m@g}%
\ifams
\F@nt \f@ntsize msa\@nd=msam10 \@step{\m@g}%
\F@nt \f@ntsize msb\@nd=msbm10 \@step{\m@g}\else\fi}
\def\smallf@nts{%
\F@nt \f@ntsize rm\@nd=cmr\s@ze
\F@nt \f@ntsize i\@nd=cmmi\s@ze
\Sk@w \f@ntsize i\@nd='177
\F@nt \f@ntsize sy\@nd=cmsy\s@ze
\Sk@w \f@ntsize sy\@nd='60
\F@nt \f@ntsize bf\@nd=cmbx\s@ze
\ifams
\F@nt \f@ntsize bdi\@nd=cmmib\s@ze
\F@nt \f@ntsize bsy\@nd=cmbsy\s@ze
\F@nt \f@ntsize msa\@nd=msam\s@ze
\F@nt \f@ntsize msb\@nd=msbm\s@ze
\else
\F@nt \f@ntsize bdi\@nd=cmmib10 \@t{\s@ze}%
\F@nt \f@ntsize bsy\@nd=cmbsy10 \@t{\s@ze}\fi
\Sk@w \f@ntsize bdi\@nd='177
\Sk@w \f@ntsize bsy\@nd='60
\F@nt \f@ntsize bss\@nd=cmssbx10 \@t{\s@ze}}%
%
%
\def\headfonts{%
\textfont0=\csname\HEAD rm\@nd
\scriptfont0=\csname\smHEAD rm\@nd
\scriptscriptfont0=\csname\vsHEAD rm\@nd
\def\rm{\fam0\csname\HEAD rm\@nd
\def\sc{\csname\smHEAD rm\@nd}}%
\textfont1=\csname\HEAD i\@nd
\scriptfont1=\csname\smHEAD i\@nd
\scriptscriptfont1=\csname\vsHEAD i\@nd
\textfont2=\csname\HEAD sy\@nd
\scriptfont2=\csname\smHEAD sy\@nd
\scriptscriptfont2=\csname\vsHEAD sy\@nd
\textfont3=\csname\HEAD ex\@nd
\scriptfont3=\csname\smHEAD ex\@nd
\scriptscriptfont3=\csname\smHEAD ex\@nd
\textfont\itfam=\csname\HEAD it\@nd
\scriptfont\itfam=\csname\smHEAD it\@nd
\scriptscriptfont\itfam=\csname\vsHEAD it\@nd
\def\it{\fam\itfam\csname\HEAD it\@nd
\def\sc{\csname\smHEAD it\@nd}}%
\textfont\slfam=\csname\HEAD sl\@nd
\def\sl{\fam\slfam\csname\HEAD sl\@nd
\def\sc{\csname\smHEAD sl\@nd}}%
\textfont\bffam=\csname\HEAD bf\@nd
\scriptfont\bffam=\csname\smHEAD bf\@nd
\scriptscriptfont\bffam=\csname\vsHEAD bf\@nd
\def\bf{\fam\bffam\csname\HEAD bf\@nd
\def\sc{\csname\smHEAD bf\@nd}}%
\textfont\ttfam=\csname\HEAD tt\@nd
\def\tt{\fam\ttfam\csname\HEAD tt\@nd}%
\textfont\bdifam=\csname\HEAD bdi\@nd
\scriptfont\bdifam=\csname\smHEAD bdi\@nd
\scriptscriptfont\bdifam=\csname\vsHEAD bdi\@nd
\def\bdi{\fam\bdifam\csname\HEAD bdi\@nd}%
\textfont\bsyfam=\csname\HEAD bsy\@nd
\scriptfont\bsyfam=\csname\smHEAD bsy\@nd
\def\bsy{\fam\bsyfam\csname\HEAD bsy\@nd}%
\textfont\bssfam=\csname\HEAD bss\@nd
\scriptfont\bssfam=\csname\smHEAD bss\@nd
\scriptscriptfont\bssfam=\csname\vsHEAD bss\@nd
\def\bss{\fam\bssfam\csname\HEAD bss\@nd}%
\ifams
\textfont\msafam=\csname\HEAD msa\@nd
\scriptfont\msafam=\csname\smHEAD msa\@nd
\scriptscriptfont\msafam=\csname\vsHEAD msa\@nd
\textfont\msbfam=\csname\HEAD msb\@nd
\scriptfont\msbfam=\csname\smHEAD msb\@nd
\scriptscriptfont\msbfam=\csname\vsHEAD msb\@nd
\else\fi
\normalbaselineskip=\headb@seline pt%
\setbox\strutbox=\hbox{\vrule height.7\normalbaselineskip
depth.3\baselineskip width0pt}%
\def\sc{\csname\smHEAD rm\@nd}\normalbaselines\bf}
%
%
\def\textfonts{%
\textfont0=\csname\TEXT rm\@nd
\scriptfont0=\csname\smTEXT rm\@nd
\scriptscriptfont0=\csname\vsTEXT rm\@nd
\def\rm{\fam0\csname\TEXT rm\@nd
\def\sc{\csname\smTEXT rm\@nd}}%
\textfont1=\csname\TEXT i\@nd
\scriptfont1=\csname\smTEXT i\@nd
\scriptscriptfont1=\csname\vsTEXT i\@nd
\textfont2=\csname\TEXT sy\@nd
\scriptfont2=\csname\smTEXT sy\@nd
\scriptscriptfont2=\csname\vsTEXT sy\@nd
\textfont3=\csname\TEXT ex\@nd
\scriptfont3=\csname\smTEXT ex\@nd
\scriptscriptfont3=\csname\smTEXT ex\@nd
\textfont\itfam=\csname\TEXT it\@nd
\scriptfont\itfam=\csname\smTEXT it\@nd
\scriptscriptfont\itfam=\csname\vsTEXT it\@nd
\def\it{\fam\itfam\csname\TEXT it\@nd
\def\sc{\csname\smTEXT it\@nd}}%
\textfont\slfam=\csname\TEXT sl\@nd
\def\sl{\fam\slfam\csname\TEXT sl\@nd
\def\sc{\csname\smTEXT sl\@nd}}%
\textfont\bffam=\csname\TEXT bf\@nd
\scriptfont\bffam=\csname\smTEXT bf\@nd
\scriptscriptfont\bffam=\csname\vsTEXT bf\@nd
\def\bf{\fam\bffam\csname\TEXT bf\@nd
\def\sc{\csname\smTEXT bf\@nd}}%
\textfont\ttfam=\csname\TEXT tt\@nd
\def\tt{\fam\ttfam\csname\TEXT tt\@nd}%
\textfont\bdifam=\csname\TEXT bdi\@nd
\scriptfont\bdifam=\csname\smTEXT bdi\@nd
\scriptscriptfont\bdifam=\csname\vsTEXT bdi\@nd
\def\bdi{\fam\bdifam\csname\TEXT bdi\@nd}%
\textfont\bsyfam=\csname\TEXT bsy\@nd
\scriptfont\bsyfam=\csname\smTEXT bsy\@nd
\def\bsy{\fam\bsyfam\csname\TEXT bsy\@nd}%
\textfont\bssfam=\csname\TEXT bss\@nd
\scriptfont\bssfam=\csname\smTEXT bss\@nd
\scriptscriptfont\bssfam=\csname\vsTEXT bss\@nd
\def\bss{\fam\bssfam\csname\TEXT bss\@nd}%
\ifams
\textfont\msafam=\csname\TEXT msa\@nd
\scriptfont\msafam=\csname\smTEXT msa\@nd
\scriptscriptfont\msafam=\csname\vsTEXT msa\@nd
\textfont\msbfam=\csname\TEXT msb\@nd
\scriptfont\msbfam=\csname\smTEXT msb\@nd
\scriptscriptfont\msbfam=\csname\vsTEXT msb\@nd
\else\fi
\normalbaselineskip=\textb@seline pt
\setbox\strutbox=\hbox{\vrule height.7\normalbaselineskip
depth.3\baselineskip width0pt}%
\everymath{}%
\def\sc{\csname\smTEXT rm\@nd}\normalbaselines\rm}
%
%
\def\smallfonts{%
\textfont0=\csname\SMALL rm\@nd
\scriptfont0=\csname\smSMALL rm\@nd
\scriptscriptfont0=\csname\vsSMALL rm\@nd
\def\rm{\fam0\csname\SMALL rm\@nd
\def\sc{\csname\smSMALL rm\@nd}}%
\textfont1=\csname\SMALL i\@nd
\scriptfont1=\csname\smSMALL i\@nd
\scriptscriptfont1=\csname\vsSMALL i\@nd
\textfont2=\csname\SMALL sy\@nd
\scriptfont2=\csname\smSMALL sy\@nd
\scriptscriptfont2=\csname\vsSMALL sy\@nd
\textfont3=\csname\SMALL ex\@nd
\scriptfont3=\csname\smSMALL ex\@nd
\scriptscriptfont3=\csname\smSMALL ex\@nd
\textfont\itfam=\csname\SMALL it\@nd
\scriptfont\itfam=\csname\smSMALL it\@nd
\scriptscriptfont\itfam=\csname\vsSMALL it\@nd
\def\it{\fam\itfam\csname\SMALL it\@nd
\def\sc{\csname\smSMALL it\@nd}}%
\textfont\slfam=\csname\SMALL sl\@nd
\def\sl{\fam\slfam\csname\SMALL sl\@nd
\def\sc{\csname\smSMALL sl\@nd}}%
\textfont\bffam=\csname\SMALL bf\@nd
\scriptfont\bffam=\csname\smSMALL bf\@nd
\scriptscriptfont\bffam=\csname\vsSMALL bf\@nd
\def\bf{\fam\bffam\csname\SMALL bf\@nd
\def\sc{\csname\smSMALL bf\@nd}}%
\textfont\ttfam=\csname\SMALL tt\@nd
\def\tt{\fam\ttfam\csname\SMALL tt\@nd}%
\textfont\bdifam=\csname\SMALL bdi\@nd
\scriptfont\bdifam=\csname\smSMALL bdi\@nd
\scriptscriptfont\bdifam=\csname\vsSMALL bdi\@nd
\def\bdi{\fam\bdifam\csname\SMALL bdi\@nd}%
\textfont\bsyfam=\csname\SMALL bsy\@nd
\scriptfont\bsyfam=\csname\smSMALL bsy\@nd
\def\bsy{\fam\bsyfam\csname\SMALL bsy\@nd}%
\textfont\bssfam=\csname\SMALL bss\@nd
\scriptfont\bssfam=\csname\smSMALL bss\@nd
\scriptscriptfont\bssfam=\csname\vsSMALL bss\@nd
\def\bss{\fam\bssfam\csname\SMALL bss\@nd}%
\ifams
\textfont\msafam=\csname\SMALL msa\@nd
\scriptfont\msafam=\csname\smSMALL msa\@nd
\scriptscriptfont\msafam=\csname\vsSMALL msa\@nd
\textfont\msbfam=\csname\SMALL msb\@nd
\scriptfont\msbfam=\csname\smSMALL msb\@nd
\scriptscriptfont\msbfam=\csname\vsSMALL msb\@nd
\else\fi
\normalbaselineskip=\smallb@seline pt%
\setbox\strutbox=\hbox{\vrule height.7\normalbaselineskip
depth.3\baselineskip width0pt}%
\everymath{}%
\def\sc{\csname\smSMALL rm\@nd}\normalbaselines\rm}%
\everydisplay{\indenteddisplay
   \gdef\labeltype{\eqlabel}}%
%
%
\def\hexnumber@#1{\ifcase#1 0\or 1\or 2\or 3\or 4\or 5\or 6\or 7\or 8\or
 9\or A\or B\or C\or D\or E\or F\fi}
\edef\bffam@{\hexnumber@\bffam}
\edef\bdifam@{\hexnumber@\bdifam}
\edef\bsyfam@{\hexnumber@\bsyfam}
\def\undefine#1{\let#1\undefined}
\def\newsymbol#1#2#3#4#5{\let\next@\relax
 \ifnum#2=\thr@@\let\next@\bdifam@\else
 \ifams
 \ifnum#2=\@ne\let\next@\msafam@\else
 \ifnum#2=\tw@\let\next@\msbfam@\fi\fi
 \fi\fi
 \mathchardef#1="#3\next@#4#5}
\def\mathhexbox@#1#2#3{\relax
 \ifmmode\mathpalette{}{\m@th\mathchar"#1#2#3}%
 \else\leavevmode\hbox{$\m@th\mathchar"#1#2#3$}\fi}

\def\bi#1{{\fam\bdifam\relax#1}}
%
%
\ifams\input amsmacro\fi
%
%
\newsymbol\bitGamma 3000
\newsymbol\bitDelta 3001
\newsymbol\bitTheta 3002
\newsymbol\bitLambda 3003
\newsymbol\bitXi 3004
\newsymbol\bitPi 3005
\newsymbol\bitSigma 3006
\newsymbol\bitUpsilon 3007
\newsymbol\bitPhi 3008
\newsymbol\bitPsi 3009
\newsymbol\bitOmega 300A
\newsymbol\balpha 300B
\newsymbol\bbeta 300C
\newsymbol\bgamma 300D
\newsymbol\bdelta 300E
\newsymbol\bepsilon 300F
\newsymbol\bzeta 3010
\newsymbol\bfeta 3011
\newsymbol\btheta 3012
\newsymbol\biota 3013
\newsymbol\bkappa 3014
\newsymbol\blambda 3015
\newsymbol\bmu 3016
\newsymbol\bnu 3017
\newsymbol\bxi 3018
\newsymbol\bpi 3019
\newsymbol\brho 301A
\newsymbol\bsigma 301B
\newsymbol\btau 301C
\newsymbol\bupsilon 301D
\newsymbol\bphi 301E
\newsymbol\bchi 301F
\newsymbol\bpsi 3020
\newsymbol\bomega 3021
\newsymbol\bvarepsilon 3022
\newsymbol\bvartheta 3023
\newsymbol\bvaromega 3024
\newsymbol\bvarrho 3025
\newsymbol\bvarzeta 3026
\newsymbol\bvarphi 3027
\newsymbol\bpartial 3040
\newsymbol\bell 3060
\newsymbol\bimath 307B
\newsymbol\bjmath 307C
\mathchardef\binfty "0\bsyfam@31
\mathchardef\bnabla "0\bsyfam@72
\mathchardef\bdot "2\bsyfam@01
\mathchardef\bGamma "0\bffam@00
\mathchardef\bDelta "0\bffam@01
\mathchardef\bTheta "0\bffam@02
\mathchardef\bLambda "0\bffam@03
\mathchardef\bXi "0\bffam@04
\mathchardef\bPi "0\bffam@05
\mathchardef\bSigma "0\bffam@06
\mathchardef\bUpsilon "0\bffam@07
\mathchardef\bPhi "0\bffam@08
\mathchardef\bPsi "0\bffam@09
\mathchardef\bOmega "0\bffam@0A
\mathchardef\itGamma "0100
\mathchardef\itDelta "0101
\mathchardef\itTheta "0102
\mathchardef\itLambda "0103
\mathchardef\itXi "0104
\mathchardef\itPi "0105
\mathchardef\itSigma "0106
\mathchardef\itUpsilon "0107
\mathchardef\itPhi "0108
\mathchardef\itPsi "0109
\mathchardef\itOmega "010A
\mathchardef\Gamma "0000
\mathchardef\Delta "0001
\mathchardef\Theta "0002
\mathchardef\Lambda "0003
\mathchardef\Xi "0004
\mathchardef\Pi "0005
\mathchardef\Sigma "0006
\mathchardef\Upsilon "0007
\mathchardef\Phi "0008
\mathchardef\Psi "0009
\mathchardef\Omega "000A
%
%
\newcount\firstpage  \firstpage=1  
\newcount\jnl                      
\newcount\secno                    
\newcount\subno                    
\newcount\subsubno                 
\newcount\appno                    
\newcount\tabno                    
\newcount\figno                    
\newcount\countno                  
\newcount\refno                    
\newcount\eqlett     \eqlett=97    
\newif\ifletter
\newif\ifwide
\newif\ifnotfull
\newif\ifaligned
\newif\ifnumbysec
\newif\ifappendix
\newif\ifnumapp
\newif\ifssf
\newif\ifppt
\newdimen\t@bwidth
\newdimen\c@pwidth
\newdimen\digitwidth                    
\newdimen\argwidth                      
\newdimen\secindent    \secindent=5pc   
\newdimen\textind    \textind=16pt      
\newdimen\tempval                       
\newskip\beforesecskip
\def\beforesecspace{\vskip\beforesecskip\relax}
\newskip\beforesubskip
\def\beforesubspace{\vskip\beforesubskip\relax}
\newskip\beforesubsubskip
\def\beforesubsubspace{\vskip\beforesubsubskip\relax}
\newskip\secskip
\def\secspace{\vskip\secskip\relax}
\newskip\subskip
\def\subspace{\vskip\subskip\relax}
\newskip\insertskip
\def\insertspace{\vskip\insertskip\relax}
\def\sp@ce{\ifx\next*\let\next=\@ssf
               \else\let\next=\@nossf\fi\next}
\def\@ssf#1{\nobreak\secspace\global\ssftrue\nobreak}
\def\@nossf{\nobreak\secspace\nobreak\noindent\ignorespaces}
\def\subsp@ce{\ifx\next*\let\next=\@sssf
               \else\let\next=\@nosssf\fi\next}
\def\@sssf#1{\nobreak\subspace\global\ssftrue\nobreak}
\def\@nosssf{\nobreak\subspace\nobreak\noindent\ignorespaces}
\beforesecskip=24pt plus12pt minus8pt
\beforesubskip=12pt plus6pt minus4pt
\beforesubsubskip=12pt plus6pt minus4pt
\secskip=12pt plus 2pt minus 2pt
\subskip=6pt plus3pt minus2pt
\insertskip=18pt plus6pt minus6pt%
\fontdimen16\tensy=2.7pt
\fontdimen17\tensy=2.7pt
%
%
\def\eqlabel{(\ifappendix\applett
               \ifnumbysec\ifnum\secno>0 \the\secno\fi.\fi
               \else\ifnumbysec\the\secno.\fi\fi\the\countno)}
\def\seclabel{\ifappendix\ifnumapp\else\applett\fi
    \ifnum\secno>0 \the\secno
    \ifnumbysec\ifnum\subno>0.\the\subno\fi\fi\fi
    \else\the\secno\fi\ifnum\subno>0.\the\subno
         \ifnum\subsubno>0.\the\subsubno\fi\fi}
\def\tablabel{\ifappendix\applett\fi\the\tabno}
\def\figlabel{\ifappendix\applett\fi\the\figno}
\def\gac{\global\advance\countno by 1}
%
%

\def\vfootnote#1{\insert\footins\bgroup
\interlinepenalty=\interfootnotelinepenalty
\splittopskip=\ht\strutbox 
\splitmaxdepth=\dp\strutbox \floatingpenalty=20000
\leftskip=0pt \rightskip=0pt \spaceskip=0pt \xspaceskip=0pt%
\noindent\smallfonts\rm #1\ \ignorespaces\footstrut\futurelet\next\fo@t}
%
%
\def\endinsert{\egroup
    \if@mid \dimen@=\ht0 \advance\dimen@ by\dp0
       \advance\dimen@ by12\p@ \advance\dimen@ by\pagetotal
       \ifdim\dimen@>\pagegoal \@midfalse\p@gefalse\fi\fi
    \if@mid \insertspace \box0 \par \ifdim\lastskip<\insertskip
    \removelastskip \penalty-200 \insertspace \fi
    \else\insert\topins{\penalty100
       \splittopskip=0pt \splitmaxdepth=\maxdimen
       \floatingpenalty=0
       \ifp@ge \dimen@=\dp0
       \vbox to\vsize{\unvbox0 \kern-\dimen@}%
       \else\box0\nobreak\insertspace\fi}\fi\endgroup}
%
%
%
\def\ind{\hbox to \secindent{\hfill}}
%
%

%
%

%
%
\def\indeqn#1{\alignedfalse\displ@y\halign{\hbox to \displaywidth
    {$\ind\@lign\displaystyle##\hfil$}\crcr #1\crcr}}
%
%
\def\indalign#1{\alignedtrue\displ@y \tabskip=0pt
  \halign to\displaywidth{\ind$\@lign\displaystyle{##}$\tabskip=0pt
    &$\@lign\displaystyle{{}##}$\hfill\tabskip=\centering
    &\llap{$\@lign\hbox{\rm##}$}\tabskip=0pt\crcr
    #1\crcr}}
\def\fl{{\hskip-\secindent}}
\def\indenteddisplay#1$${\indispl@y{#1 }}
\def\indispl@y#1{\disptest#1\eqalignno\eqalignno\disptest}
\def\disptest#1\eqalignno#2\eqalignno#3\disptest{%
    \ifx#3\eqalignno
    \indalign#2%
    \else\indeqn{#1}\fi$$}
%
%

%
%

%
%

%
%

%
%

\def\ns{\noalign{\vskip-3pt}}

%

%
%
\def\bhbar{\rlap{\kern1pt\raise.4ex\hbox{\bf\char'40}}\bi{h}}

\def\dash{---{}--- }

\def\etal{{\it et al\/}\ }
\def\frac#1#2{{#1\over#2}}
\ifams
\def\lap{\lesssim}
\def\gap{\gtrsim}

\let\leq=\leqslant

\let\geq=\geqslant
\else

\def\gap{\;\lower3pt\hbox{$\buildrel > \over \sim$}\;}%
\def\lap{\;\lower3pt\hbox{$\buildrel < \over \sim$}\;}\fi

\def\tqs{\hbox to 25pt{\hfil}}

\def\Bbbone{1\kern-.22em {\rm l}}
%
%
\def\rp{\raise8pt\hbox{$\scriptstyle\prime$}}
%
%
%
%

%
%
\def\[#1\]{\setbox0=\hbox{$\dsty#1$}\argwidth=\wd0
    \setbox0=\hbox{$\left[\box0\right]$}\advance\argwidth by -\wd0
    \left[\kern.3\argwidth\box0\kern.3\argwidth\right]}
%
%
\def\lsb#1\rsb{\setbox0=\hbox{$#1$}\argwidth=\wd0
    \setbox0=\hbox{$\left[\box0\right]$}\advance\argwidth by -\wd0
    \left[\kern.3\argwidth\box0\kern.3\argwidth\right]}
%

%
%

%
\def\pt(#1){({\it #1\/})}
\let\dsty=\displaystyle

%
%
\def\reactions#1{\vskip 12pt plus2pt minus2pt%
\vbox{\hbox{\kern\secindent\vrule\kern12pt%
\vbox{\kern0.5pt\vbox{\hsize=24pc\parindent=0pt\smallfonts\rm NUCLEAR
REACTIONS\strut\quad #1\strut}\kern0.5pt}\kern12pt\vrule}}}
%
%
\def\slashchar#1{\setbox0=\hbox{$#1$}\dimen0=\wd0%
\setbox1=\hbox{/}\dimen1=\wd1%
\ifdim\dimen0>\dimen1%
\rlap{\hbox to \dimen0{\hfil/\hfil}}#1\else
\rlap{\hbox to \dimen1{\hfil$#1$\hfil}}/\fi}
%
%
\def\textindent#1{\noindent\hbox to \parindent{#1\hss}\ignorespaces}
%
%
\def\opencirc{\raise1pt\hbox{$\scriptstyle{\bigcirc}$}}

\ifams
\def\opensqr{\hbox{$\square$}}

\def\opentridown{\hbox{$\triangledown$}}

\else
\def\opensqr{\vbox{\hrule height.4pt\hbox{\vrule width.4pt height3.5pt
    \kern3.5pt\vrule width.4pt}\hrule height.4pt}}

\def\opentridown{\raise1pt\hbox{$\scriptstyle\bigtriangledown$}}

\fi

%
%
\def\m@th{\mathsurround=0pt}
%
%
\def\cases#1{%
\left\{\,\vcenter{\normalbaselines\openup1\jot\m@th%
     \ialign{$\displaystyle##\hfil$&\rm\tqs##\hfil\crcr#1\crcr}}\right.}%
%
%
\def\oldcases#1{\left\{\,\vcenter{\normalbaselines\m@th
    \ialign{$##\hfil$&\rm\quad##\hfil\crcr#1\crcr}}\right.}
%
%
\def\numcases#1{\left\{\,\vcenter{\baselineskip=15pt\m@th%
     \ialign{$\displaystyle##\hfil$&\rm\tqs##\hfil
     \crcr#1\crcr}}\right.\hfill
     \vcenter{\baselineskip=15pt\m@th%
     \ialign{\rlap{$\phantom{\displaystyle##\hfil}$}\tabskip=0pt&\en
     \rlap{\phantom{##\hfil}}\crcr#1\crcr}}}
\def\ptnumcases#1{\left\{\,\vcenter{\baselineskip=15pt\m@th%
     \ialign{$\displaystyle##\hfil$&\rm\tqs##\hfil
     \crcr#1\crcr}}\right.\hfill
     \vcenter{\baselineskip=15pt\m@th%
     \ialign{\rlap{$\phantom{\displaystyle##\hfil}$}\tabskip=0pt&\enpt
     \rlap{\phantom{##\hfil}}\crcr#1\crcr}}\global\eqlett=97
     \global\advance\countno by 1}
%
%
\def\eq(#1){\ifaligned\@mp(#1)\else\hfill\llap{{\rm (#1)}}\fi}
\def\ceq(#1){\ns\ns\ifaligned\@mp\fi\eq(#1)\cr\ns\ns}
\def\eqpt(#1#2){\ifaligned\@mp(#1{\it #2\/})
                    \else\hfill\llap{{\rm (#1{\it #2\/})}}\fi}
\let\eqno=\eq
%
%
\countno=1

\def\aleq{&\rm(\ifappendix\applett
               \ifnumbysec\ifnum\secno>0 \the\secno\fi.\fi
               \else\ifnumbysec\the\secno.\fi\fi\the\countno}
\def\noaleq{\hfill\llap\bgroup\rm(\ifappendix\applett
               \ifnumbysec\ifnum\secno>0 \the\secno\fi.\fi
               \else\ifnumbysec\the\secno.\fi\fi\the\countno}
\def\@mp{&}
\def\en{\ifaligned\aleq)\else\noaleq)\egroup\fi\gac}
\def\cen{\ns\ns\ifaligned\@mp\fi\en\cr\ns\ns}
\def\enpt{\ifaligned\aleq{\it\char\the\eqlett})\else
    \noaleq{\it\char\the\eqlett})\egroup\fi
    \global\advance\eqlett by 1}
\def\endpt{\ifaligned\aleq{\it\char\the\eqlett})\else
    \noaleq{\it\char\the\eqlett})\egroup\fi
    \global\eqlett=97\gac}
%
%

\def\JPA{{\it J. Phys. A: Math. Gen.}}
\def\JPC{{\it J. Phys. C: Solid State Phys.}}     



%
%

\def\APNY{{\it Ann. Phys., NY\/}}

\def\JMP{{\it J. Math. Phys.}}

\def\JPSJ{{\it J. Phys. Soc. Japan\/}}

\def\PL{{\it Phys. Lett.}}
\def\PR{{\it Phys. Rev.}}
\def\PRL{{\it Phys. Rev. Lett.}}

\def\RMP{{\it Rev. Mod. Phys.}}

\def\ZP{{\it Z. Phys.}}
\headline={\ifodd\pageno{\ifnum\pageno=\firstpage\hfill
   \else\rrhead\fi}\else\lrhead\fi}
\def\rrhead{\textfonts\hskip\secindent\it
    \shorttitle\hfill\rm\folio}
\def\lrhead{\textfonts\hbox to\secindent{\rm\folio\hss}%
    \it\aunames\hss}
\footline={\ifnum\pageno=\firstpage \hfill\textfonts\rm\folio\fi}
\def\@rticle#1#2{\vglue.5pc
    {\parindent=\secindent \bf #1\par}
     \vskip2.5pc
    {\exhyphenpenalty=10000\hyphenpenalty=10000
     \baselineskip=18pt\raggedright\noindent
     \headfonts\bf#2\par}\futurelet\next\sh@rttitle}%
\def\title#1{\gdef\shorttitle{#1}
    \vglue4pc{\exhyphenpenalty=10000\hyphenpenalty=10000
    \baselineskip=18pt
    \raggedright\parindent=0pt
    \headfonts\bf#1\par}\futurelet\next\sh@rttitle}

\def\article#1#2{\gdef\shorttitle{#2}\@rticle{#1}{#2}}
\def\review#1{\gdef\shorttitle{#1}%
    \@rticle{REVIEW \ifpbm\else ARTICLE\fi}{#1}}
\def\topical#1{\gdef\shorttitle{#1}%
    \@rticle{TOPICAL REVIEW}{#1}}
\def\comment#1{\gdef\shorttitle{#1}%
    \@rticle{COMMENT}{#1}}
\def\note#1{\gdef\shorttitle{#1}%
    \@rticle{NOTE}{#1}}
\def\prelim#1{\gdef\shorttitle{#1}%
    \@rticle{PRELIMINARY COMMUNICATION}{#1}}
\def\letter#1{\gdef\shorttitle{Letter to the Editor}%
     \gdef\aunames{Letter to the Editor}
     \global\lettertrue\ifnum\jnl=7\global\letterfalse\fi
     \@rticle{LETTER TO THE EDITOR}{#1}}
\def\sh@rttitle{\ifx\next[\let\next=\sh@rt
                \else\let\next=\f@ll\fi\next}
\def\sh@rt[#1]{\gdef\shorttitle{#1}}
\def\f@ll{}
\def\author#1{\ifletter\else\gdef\aunames{#1}\fi\vskip1.5pc
    {\parindent=\secindent
     \hang\textfonts
     \ifppt\bf\else\rm\fi#1\par}
     \ifppt\bigskip\else\smallskip\fi
     \futurelet\next\@unames}
\def\@unames{\ifx\next[\let\next=\short@uthor
                 \else\let\next=\@uthor\fi\next}
\def\short@uthor[#1]{\gdef\aunames{#1}}
\def\@uthor{}
\def\address#1{{\parindent=\secindent
    \exhyphenpenalty=10000\hyphenpenalty=10000
\ifppt\textfonts\else\smallfonts\fi\hang\raggedright\rm#1\par}%
    \ifppt\bigskip\fi}
\def\jl#1{\global\jnl=#1}
\jl{0}%
\def\journal{\ifnum\jnl=1 J. Phys.\ A: Math.\ Gen.\
        \else\ifnum\jnl=2 J. Phys.\ B: At.\ Mol.\ Opt.\ Phys.\
        \else\ifnum\jnl=3 J. Phys.:\ Condens. Matter\
        \else\ifnum\jnl=4 J. Phys.\ G: Nucl.\ Part.\ Phys.\
        \else\ifnum\jnl=5 Inverse Problems\
        \else\ifnum\jnl=6 Class. Quantum Grav.\
        \else\ifnum\jnl=7 Network\
        \else\ifnum\jnl=8 Nonlinearity\
        \else\ifnum\jnl=9 Quantum Opt.\
        \else\ifnum\jnl=10 Waves in Random Media\
        \else\ifnum\jnl=11 Pure Appl. Opt.\
        \else\ifnum\jnl=12 Phys. Med. Biol.\
        \else\ifnum\jnl=13 Modelling Simulation Mater.\ Sci.\ Eng.\
        \else\ifnum\jnl=14 Plasma Phys. Control. Fusion\
        \else\ifnum\jnl=15 Physiol. Meas.\
        \else\ifnum\jnl=16 Sov.\ Lightwave Commun.\
        \else\ifnum\jnl=17 J. Phys.\ D: Appl.\ Phys.\
        \else\ifnum\jnl=18 Supercond.\ Sci.\ Technol.\
        \else\ifnum\jnl=19 Semicond.\ Sci.\ Technol.\
        \else\ifnum\jnl=20 Nanotechnology\
        \else\ifnum\jnl=21 Meas.\ Sci.\ Technol.\
        \else\ifnum\jnl=22 Plasma Sources Sci.\ Technol.\
        \else\ifnum\jnl=23 Smart Mater.\ Struct.\
        \else\ifnum\jnl=24 J.\ Micromech.\ Microeng.\
   \else Institute of Physics Publishing\
   \fi\fi\fi\fi\fi\fi\fi\fi\fi\fi\fi\fi\fi\fi\fi
   \fi\fi\fi\fi\fi\fi\fi\fi\fi}
\let\abs=\beginabstract

\let\endabs=\endabstract
\def\submitted{\ifppt\noindent\textfonts\rm Submitted to \journal\par
     \bigskip\fi}
\def\today{\number\day\ \ifcase\month\or
     January\or February\or March\or April\or May\or June\or
     July\or August\or September\or October\or November\or
     December\fi\space \number\year}
\def\date{\ifppt\noindent\textfonts\rm
     Date: \today\par\goodbreak\bigskip\fi}
%
%
\def\pacs#1{\ifppt\noindent\textfonts\rm
     PACS number(s): #1\par\bigskip\fi}
%

%
%
\def\section#1{\ifppt\ifnum\secno=0\eject\fi\fi
    \subno=0\subsubno=0\global\advance\secno by 1
    \gdef\labeltype{\seclabel}\ifnumbysec\countno=1\fi
    \goodbreak\beforesecspace\nobreak
    \noindent{\bf \the\secno. #1}\par\futurelet\next\sp@ce}
\def\subsection#1{\subsubno=0\global\advance\subno by 1
     \gdef\labeltype{\seclabel}%
     \ifssf\else\goodbreak\beforesubspace\fi
     \global\ssffalse\nobreak
     \noindent{\it \the\secno.\the\subno. #1\par}%
     \futurelet\next\subsp@ce}
\def\subsubsection#1{\global\advance\subsubno by 1
     \gdef\labeltype{\seclabel}%
     \ifssf\else\goodbreak\beforesubsubspace\fi
     \global\ssffalse\nobreak
     \noindent{\it \the\secno.\the\subno.\the\subsubno. #1}\null.
     \ignorespaces}
%

%
%
\def\numappendix#1{\ifappendix\ifnumbysec\countno=1\fi\else
    \countno=1\figno=0\tabno=0\fi
    \subno=0\global\advance\appno by 1
    \secno=\appno\gdef\applett{A}\gdef\labeltype{\seclabel}%
    \global\appendixtrue\global\numapptrue
    \goodbreak\beforesecspace\nobreak
    \noindent{\bf Appendix \the\appno. #1\par}%
    \futurelet\next\sp@ce}
\def\numsubappendix#1{\global\advance\subno by 1\subsubno=0
    \gdef\labeltype{\seclabel}%
    \ifssf\else\goodbreak\beforesubspace\fi
    \global\ssffalse\nobreak
    \noindent{\it A\the\appno.\the\subno. #1\par}%
    \futurelet\next\subsp@ce}
\def\@ppendix#1#2#3{\countno=1\subno=0\subsubno=0\secno=0\figno=0\tabno=0
    \gdef\applett{#1}\gdef\labeltype{\seclabel}\global\appendixtrue
    \goodbreak\beforesecspace\nobreak
    \noindent{\bf Appendix#2#3\par}\futurelet\next\sp@ce}
\def\Appendix#1{\@ppendix{A}{. }{#1}}
\def\appendix#1#2{\@ppendix{#1}{ #1. }{#2}}
\def\App#1{\@ppendix{A}{ }{#1}}
\def\app{\@ppendix{A}{}{}}
\def\subappendix#1#2{\global\advance\subno by 1\subsubno=0
    \gdef\labeltype{\seclabel}%
    \ifssf\else\goodbreak\beforesubspace\fi
    \global\ssffalse\nobreak
    \noindent{\it #1\the\subno. #2\par}%
    \nobreak\subspace\noindent\ignorespaces}
%
%
\def\@ck#1{\ifletter\bigskip\noindent\ignorespaces\else
    \goodbreak\beforesecspace\nobreak
    \noindent{\bf Acknowledgment#1\par}%
    \nobreak\secspace\noindent\ignorespaces\fi}
\def\ack{\@ck{s}}
\def\ackn{\@ck{}}
\def\n@ip#1{\goodbreak\beforesecspace\nobreak
    \noindent\smallfonts{\it #1}. \rm\ignorespaces}
\def\naip{\n@ip{Note added in proof}}
\def\na{\n@ip{Note added}}

%
%

%

%
%

%

%

\def\tablecont{\topinsert\global\advance\tabno by -1
    \tablecaption{(continued)}}
\def\tablecaption#1{\gdef\labeltype{\tablabel}\global\widefalse
    \leftskip=\secindent\parindent=0pt
    \global\advance\tabno by 1
    \smallfonts{\bf Table \ifappendix\applett\fi\the\tabno.} \rm #1\par
    \smallskip\futurelet\next\t@b}
\def\t@b{\ifx\next*\let\next=\widet@b
             \else\ifx\next[\let\next=\fullwidet@b
                      \else\let\next=\narrowt@b\fi\fi
             \next}
\def\widet@b#1{\global\widetrue\global\notfulltrue
    \t@bwidth=\hsize\advance\t@bwidth by -\secindent}
\def\fullwidet@b[#1]{\global\widetrue\global\notfullfalse
    \leftskip=0pt\t@bwidth=\hsize}
\def\narrowt@b{\global\notfulltrue}
\def\align{\catcode`?=13\ifnotfull\moveright\secindent\fi
    \vbox\bgroup\halign\ifwide to \t@bwidth\fi
    \bgroup\strut\tabskip=1.2pc plus1pc minus.5pc}
\def\endalign{\egroup\egroup\catcode`?=12}

%
%

%
%

%

%
%

%

\catcode`?=13
\def\lineup{\setbox0=\hbox{\smallfonts\rm 0}%
    \digitwidth=\wd0%
    \def?{\kern\digitwidth}%
    \def\\{\hbox{$\phantom{-}$}}%
    \def\-{\llap{$-$}}}
\catcode`?=12
%
%
\def\sidetable#1#2{\hbox{\ifppt\hsize=18pc\t@bwidth=18pc
                          \else\hsize=15pc\t@bwidth=15pc\fi
    \parindent=0pt\vtop{\null #1\par}%
    \ifppt\hskip1.2pc\else\hskip1pc\fi
    \vtop{\null #2\par}}}
\def\lstable#1#2{\everypar{}\tempval=\hsize\hsize=\vsize
    \vsize=\tempval\hoffset=-3pc
    \global\tabno=#1\gdef\labeltype{\tablabel}%
    \noindent\smallfonts{\bf Table \ifappendix\applett\fi
    \the\tabno.} \rm #2\par
    \smallskip\futurelet\next\t@b}
\def\inctabno{\global\advance\tabno by 1}
%
%

%

%
\def\figure#1{\figc@ption{#1}\bigskip}
\def\figc@ption#1{\global\advance\figno by 1\gdef\labeltype{\figlabel}%
   {\parindent=\secindent\smallfonts\hang
    {\bf Figure \ifappendix\applett\fi\the\figno.} \rm #1\par}}
%
%
\def\refHEAD{\goodbreak\beforesecspace
     \noindent\textfonts{\bf References}\par
     \let\ref=\rf
     \nobreak\smallfonts\rm}
\def\references{\refHEAD\parindent=0pt
     \everypar{\hangindent=18pt\hangafter=1
     \frenchspacing\rm}%
     \secspace}
\def\rf#1{\par\noindent\hbox to 21pt{\hss #1\quad}\ignorespaces}
\def\refjl#1#2#3#4{\noindent #1 {\it #2 \bf #3} #4\par}
\def\refbk#1#2#3{\noindent #1 {\it #2} #3\par}
%
%

%
%

%
%

%
%

%
\catcode`\@=12
%
%

%
%
\def\jnlstyle{\pptfalse\headsize{14}{18}%
\textsize{10}{12}%
\smallsize{8}{10}
\textind=16pt}
%
%

%
%

%
\parindent=\textind
%
\input epsf

\def\figure#1{\global\advance\figno by 1\gdef\labeltype{\figlabel}%
   {\parindent=\secindent\smallfonts\hang
    {\bf Figure \ifappendix\applett\fi\the\figno.} \rm #1\par}}
\headline={\ifodd\pageno{\ifnum\pageno=\firstpage\titlehead
   \else\rrhead\fi}\else\lrhead\fi}
\def\lpsn#1#2{LPSN-#1-LT#2}

\footline={\ifnum\pageno=\firstpage{\smallfonts cond--mat/9411030}
\hfil\textfonts\rm\folio\fi}
\def\titlehead{\smallfonts J. Phys. A: Math. Gen.  {\bf 28} (1995) 45--55
\hfil\lpsn{94}{4}}

\firstpage=45
\pageno=45

\jnlstyle
\jl{1}
\overfullrule=0pt

\title{Local critical behaviour at aperiodic surface extended perturbation in
the Ising quantum chain}[Aperiodic surface extended perturbation]

\author{Dragi Karevski, G\'abor Pal\'agyi\footnote\dag{\eightrm Permanent
address: Department of Theoretical Physics, University of Szeged, H--6720
Szeged, Hungary } and Lo\"\i c Turban}[D Karevski \etal]

\address{Laboratoire de Physique du Solide\footnote\ddag{\eightrm Unit\'e de
Recherche Associ\'ee au CNRS No 155},  Universit\'e Henri Poincar\'e (Nancy~I),
BP 239, F--54506 Vand\oe uvre l\`es Nancy Cedex, France}

\abs
The surface critical behaviour of the semi--infinite one--dimensional quantum
Ising model in a transverse field  is studied in the presence of an
aperiodic surface extended modulation. The perturbed
couplings are distributed according to a generalized Fredholm
sequence, leading to a marginal perturbation and varying surface exponents. The
surface magnetic exponents are calculated exactly whereas the expression
of the surface energy density exponent is conjectured from a finite--size
scaling study. The system displays surface order at the bulk critical point,
above a critical value of the modulation amplitude. It may be considered as a
discrete realization of the Hilhorst--van Leeuwen model.
\endabs

\pacs{05.50.+q, 64.60.Cn, 64.60.Fr}
\submitted
\date

\section{Introduction}
The influence of bulk quasiperiodic or aperiodic perturbations on the critical
properties at second order phase transitions has been an active
field of research during the last years. Different problems were
studied numerically on the two--dimensional Penrose lattice
including the Ising model (Godr\`eche \etal 1986, Okabe and Niizeki
1988, S\o rensen \etal 1991), percolation (Sakamoto
\etal 1989, Zhang and De'Bell 1993) and the self--avoiding--walk
(Langie and Igl\'oi 1992). In all these cases, no change in the critical
exponents was observed. Universality was also preserved for
three--dimensional quasiperiodic systems (Okabe and Niizeki 1990). On the
contrary, a continuously varying roughness exponent was obtained for interface
roughening in two dimensions with a modulation of the couplings following the
Fibonacci sequence (Henley and Lipowsky 1987, Garg and Levine 1987).

The aperiodically layered two-dimensional Ising model has been also extensively
studied (Igl\'oi 1988, Doria and Satija 1988, Benza 1989, Ceccato 1989a, 1989b,
Henkel and Patk\'os 1992, Lin and Tao 1990,1992a,1992b, You \etal 1992, Turban
and Berche 1993) following earlier pioneering works on randomly or
arbitrarily layered Ising systems (McCoy and Wu 1968, Au--Yang and McCoy 1974).
The problem was mainly treated in the extreme anisotropic limit where the
constant intralayer interaction $K_1\!=\!J_1/k_BT$ goes to infinity while the
modulated interlayer interactions $K_2(k)$ go to zero, keeping fixed the
ratio $\lambda_k\!=\! K_2(k)/K_1^*$ where $K_1^*$ is related to~$K_1$ through
duality (Kogut 1979). In this limit, the physics of the system is governed by
the one--dimensional quantum Ising model (QIM) with the Hamiltonian
$$
{\cal H}=-{1\over 2}\sum_{k=1}^\infty\big[\sigma_k^z+\lambda_k\sigma_k^x
\sigma_{k+1}^x\big] .
\eqno(1.1)
$$
Some exact results have been obtained with Fibonacci, Thue--Morse and other
aperiodic modulations for which the critical behaviour is universal. In other
cases the Onsager logarithmic singularity of the specific heat was found to be
washed out (Tracy 1988), like in the random McCoy--Wu model. The structure of
the critical excitation spectrum and the related conformal aspects have been
also explored (Igl\'oi 1988, Grimm and Baake 1994).

The situation was recently clarified through the introduction of a
relevance--irrelevance criterion (Luck 1993a, 1993b, Igl\'oi 1993)
generalizing to aperiodic systems the Harris criterion for random systems
(Harris 1974).
The cumulated deviation from the average coupling $\overline{\lambda}$,
$\Delta(L)\!=\!\sum_{k\!=\!1}^L\big(\lambda_k\!-\!\overline{\lambda}\big)$,
scales with the chain length $L$ as $\delta L^\omega$ where $\delta$ is the
amplitude of the modulation and $\omega$ is the wandering exponent of the
aperiodic sequence related to the leading eigenvalues of its
substitution matrix (Queffelec 1987, Dumont 1990). Under a change of the
length scale by a factor $b\!=\! L/L'$, the average thermal perturbation
$\Delta(L)/L$ is multiplied by $b^{1/\nu}$, where $\nu$ is the bulk correlation
length exponent, and the amplitude transforms as
$$
\delta'=b^{\Phi/\nu}\delta ,\qquad \Phi=1+\nu(\omega-1) .
\eqno(1.2)
$$
The relevance of the aperiodic perturbation depends on the sign of the
crossover exponent $\Phi$. In the two--dimensional Ising model with
$\nu\!=\!1$ the modulation is irrelevant when $\omega\!<\!0$, i. e. for the
much studied Fibonacci and Thue--Morse sequences, marginal when $\omega\!=\!0$
and relevant when $\omega\!>\!0$. One expects varying exponents In the
marginal case and a new type of critical behaviour in the relevant case.

The three types of critical behaviour were indeed obtained in recent exact
calculations of the surface magnetization of the QIM (Turban, Igl\'oi and
Berche 1994, Igl\'oi and Turban 1994, Turban, Berche and Berche 1994).

In the present work, we study surface aperiodic perturbations
generated through the Fredhom sequence (Dekking \etal 1983) and its
generalizations. Such sequences lead to a vanishing density of defects in the
bulk of the system and the bulk critical properties are left unchanged. On
the other hand, they induce a marginal surface extended perturbation so that
the
surface critical exponents are nonuniversal, varying with the modulation
amplitude. Such aperiodic perturbations may be considered as discrete
realizations of the Hilhorst--van Leeuwen model (Hilhorst and van Leeuwen
1981, see also Igl\'oi, Peschel and Turban 1993, for a recent review).

The properties of the generalized Fredholm sequence are studied in section 2.
The surface magnetization is calculated exactly in section 3 and the surface
energy exponent is obtained through finite--size scaling in section 4. The
results are discussed in the last section.

\section{Generalized Fredholm sequence}
We consider a generalized Fredholm sequence generated through substitution on
the three letters $A$, $B$ and $C$
$$
\eqalign{
A\to {\cal S}(A)&=A\  B\  C\ C\  \cdots\  C\cr
B\to {\cal S}(B)&=B\  C\  C\ C\  \cdots\  C\cr
C\to {\cal S}(C)&=\underbrace{C\  C\  C\ C\  \cdots\  C}_{m}\cr}
\eqno(2.1)
$$
which is the characteristic sequence of the powers of $m$.
With words of length $m\!=\!2$, one recovers the usual Fredholm sequence
(Dekking \etal 1983).

The substitution matrix $\bss{M}$, with entries giving the numbers of $A$, $B$
and $C$ in ${\cal S}(A)$, ${\cal S}(B)$ and ${\cal S}(C)$, then reads
$$
\bss{M}=\left(\matrix{1&0&0\cr 1&1&0\cr m-2&m-1&m\cr}\right),
\eqno(2.2)
$$
with eigenvalues $\Omega_1\!=\! m$, $\Omega_{2,3}\!=\!1$. The wandering
exponent
is given by
$$
\omega={\ln\Omega_2\over\ln\Omega_1}=0
\eqno(2.3)
$$
so that, according to equation (1.2), the Fredholm modulation is a marginal
perturbation for the QIM.
 The numbers of letters of each type at
the $n$th step in the inflation process are given by the matrix elements of
$\bss{M}^n$. The asymptotic letter densities $\rho_A$, $\rho_B$ and $\rho_C$
are related to the components of the right eigenvector associated with the
largest eigenvalue $\Omega_1$ with here
$$
\rho_A=0,\qquad\rho_B=0,\qquad\rho_C=1,
\eqno(2.4)
$$
independent of $m$.

In the following, the couplings in the Hamiltonian (1.1), which are distributed
according to the aperiodic sequence, will be written as $\lambda_k\!=\lambda
r^{f_k}$ where $\lambda$ is the unperturbed (bulk) interaction and $r$
characterizes the modulation. We associate an unperturbed coupling $\lambda$,
i.
e. $f_k\!=\!0$, to the letters $A$ or $C$ and a perturbed coupling $\lambda r$,
$f_k\!=\!1$, to $B$. For example, starting on $A$ with $m\!=\!2$, one obtains
the following sequences after $n$ iterations: $$
\fl\matrix{
n=0\quad&A&&&&&&&&&&&&&&&\cr
n=1\quad&A&B&&&&&&&&&&&&&&\cr
n=2\quad&A&B&B&C&&&&&&&&&&&&\cr
n=3\quad&A&B&B&C&B&C&C&C&&&&&&&&\cr
n=4\quad&A&B&B&C&B&C&C&C&B&C&C&C&C&C&C&C\cr
f_k\quad&0&1&1&0&1&0&0&0&1&0&0&0&0&0&0&0\cr}
\eqno(2.5)
$$
Equation (2.1) leads to the following relations for the $f_k$s:
$$
\eqalign{
f_{mp+1}=&f_{p+1} ,\cr
f_{mp+2}=&0\qquad (p>0) ,\qquad f_2=1 ,\cr
f_{mp+q}=&0\qquad (q=3,4,\dots,m) .\cr}
\eqno(2.6)
$$
They can be used to deduce similar recursion relations for the number of
perturbed couplings $n_j=\sum_{k=1}^jf_k$ in a sequence with length $j$, which
are obtained by splitting the sum over $k$ into $m$ sums over $p$, giving:
$$
\eqalign{
n_{ml+1}=&n_{l+1}+1\qquad (l>0) ,\qquad n_1=0 ,\cr
n_{ml+q}=&n_{l+1}+1\qquad (q=2,3,\dots,m) .\cr}
\eqno(2.7)
$$
Iterating these relations, one may check that $f_k\!=\!1$ when
$k\!=\!m^l\!+\!1$ $(l=0,1,2,\dots)$. For a sequence with length $L\!=\!m^l$,
$$
n_L=l={\ln L\over\ln m} ,
\eqno(2.8)
$$
and the asymptotic density of defects satisfies
$$
\rho_\infty=\lim_{L\to\infty}{n_L\over L}=\rho_B=0 ,
\eqno(2.9)
$$
i. e. the generalized Fredholm modulation introduces an extended surface
per\-tur\-ba\-tion in the system.

\section{Surface magnetization}
The surface magnetization $m_s$ follows from the asymptotic behaviour of the
surface spin--spin correlation function,
$\lim_{t\to\infty}\langle\sigma_1^x(0)\sigma_1^x(t)\rangle$, which gives the
square of this quantity on a semi--infinite system. Writing the correlation
function in the basis which diagonalizes the Hamiltonian in equation (1.1),
$m_s$ can be expressed as the matrix element
$\langle\sigma\!\mid\!\sigma^x\!\mid\!0\rangle$, where $\!\mid\!0\rangle$ is
the
groundstate and $\!\mid\!\sigma\rangle$ the first excited state of $\cal{H}$
(Schultz~\etal~1964). These two states become degenerate in the ordered phase
$\lambda>\lambda_c$ as a consequence of long--range order.

The Hamiltonian can be put in diagonal form (Lieb \etal 1961)
$$
{\cal H}=\sum_\nu\epsilon_\nu\left(\eta_\nu^\dagger\eta_\nu-{1\over
2}\right)
\eqno(3.1)
$$
using the Jordan--Wigner transformation (Jordan~and~Wigner~1928) followed by
a canonical transformation to the diagonal fermion operators $\eta_\nu$.
The fermion excitation spectrum is obtained as the solution of the
eigenvalue problem
$$
\eqalign{
\epsilon_\nu\psi_\nu(k)&=-\phi_\nu(k)-\lambda_k
\phi_\nu(k+1)\cr
\epsilon_\nu\phi_\nu(k)&=-\lambda_{k-1}\psi_\nu(k-1)-\psi_\nu(k)\cr}
\eqno(3.2)
$$
where the $\phi_\nu(k)$ and $\psi_\nu(k)$ are the components of
two normalized eigenvectors which satisfy the boundary conditions
$\phi_\nu(0)\!=\!\psi_\nu(0)\!=\! 0$.

Rewriting $\sigma^x$ in terms of diagonal fermions with
$\!\mid\!\sigma\rangle\!=\!\eta_1^\dagger\!\mid\!0\rangle$, it can be shown
that
$m_s$ is also given by the first component $\phi_1(1)$ of the eigenvector
corresponding to the smallest excitation.
According to the first equation in~(3.2), in the ordered phase where
$\epsilon_1$ vanishes, other components of the eigenvector follow from the
recursion relation
$$
\phi_{1}(k+1)=-\lambda_k^{-1}\phi_1(k).
\eqno(3.3)
$$
The normalization of the eigenvector then leads to the surface magnetization
(Peschel~1984)
$$
m_s=\left(1+\sum_{j=1}^\infty\prod_{k=1}^j\lambda_k^{-2}\right)^{-1/2}.
\eqno(3.4)
$$
For the aperiodic system, with $\lambda_k\!=\!\lambda r^{f_k}$, this leads to:
$$
m_s=\left[S(\lambda,r)\right]^{-1/2} ,\qquad S(\lambda,r)=\sum_{j=0}^\infty
\lambda^{-2j}r^{-2n_j} ,\qquad n_0=0 .
\eqno(3.5)
$$
The critical coupling $\lambda_c$ generally follows from (Pfeuty~1979):
$$
\lim_{j\to\infty}{1/j}\sum_{k=1}^j\ln\lambda_k\!=\!0 .
\eqno(3.6)
$$
Here, $\lambda_c\!=\! r^{-\rho_\infty}\!=\!1$ keeps its
unperturbed value, as expected for an extended surface perturbation.

In order to calculate the sum in (3.5) let us rewrite it as
$$
S(\lambda,r)=1+\lambda^{-2}+T(\lambda,r) ,\qquad
T(\lambda,r)=\sum_{j=2}^\infty\lambda^{-2j}r^{-2n_j} .
\eqno(3.7)
$$
The second sum can be splitted into $m$ parts as
$$\eqalign{
T(\lambda,r)=&\sum_{l=1}^\infty\lambda^{-2(ml+1)}r^{-2n_{ml+1}}+
\sum_{l=0}^\infty\lambda^{-2(ml+2)}r^{-2n_{ml+2}}+\cdots\cr
&+\sum_{l=0}^\infty\lambda^{-2(ml+m)}r^{-2n_{ml+m}}\cr}
\eq(3.8)
$$
and, using (2.7), the following functional equation is obtained:
$$
T(\lambda,r)={r^{-2}\over \lambda^2-1}\left[\lambda^{-2}-\lambda^{-2m}+
(\lambda^{2m}-1)\ T(\lambda^m,r)\right].
\eqno(3.9)
$$
This can be iterated to give:
$$
T(\lambda,r)=\sum_{l=0}^\infty
r^{-2(l+1)}\lambda^{-2m^{l+1}}\sum_{p=0}^{(m-1)m^l-1}\lambda^{2p} .
\eqno(3.10)
$$

The critical behaviour can be extracted, applying a finite-size scaling method
due to Igl\'oi. Assuming that the  surface magnetization displays a power--law
singularity with a critical exponent $\beta_s$, $S(\lambda,r)$ behaves as
$t^{-2\beta_s}$ with $t\!=\! \lambda_c^{-2}-\lambda^{-2}$,
near the critical point $\lambda_c\!=\!1$. It can be shown (Igl\'oi~1986) that
the sum $S_L$ of the first $L$ terms in a power series expansion in
$\lambda^{-2}$ asymptotically scales like $L^{2\beta_s}$ at the critical point.

{\par\begingroup\parindent=0pt\medskip
\epsfxsize=9truecm
\topinsert
\centerline{\epsfbox{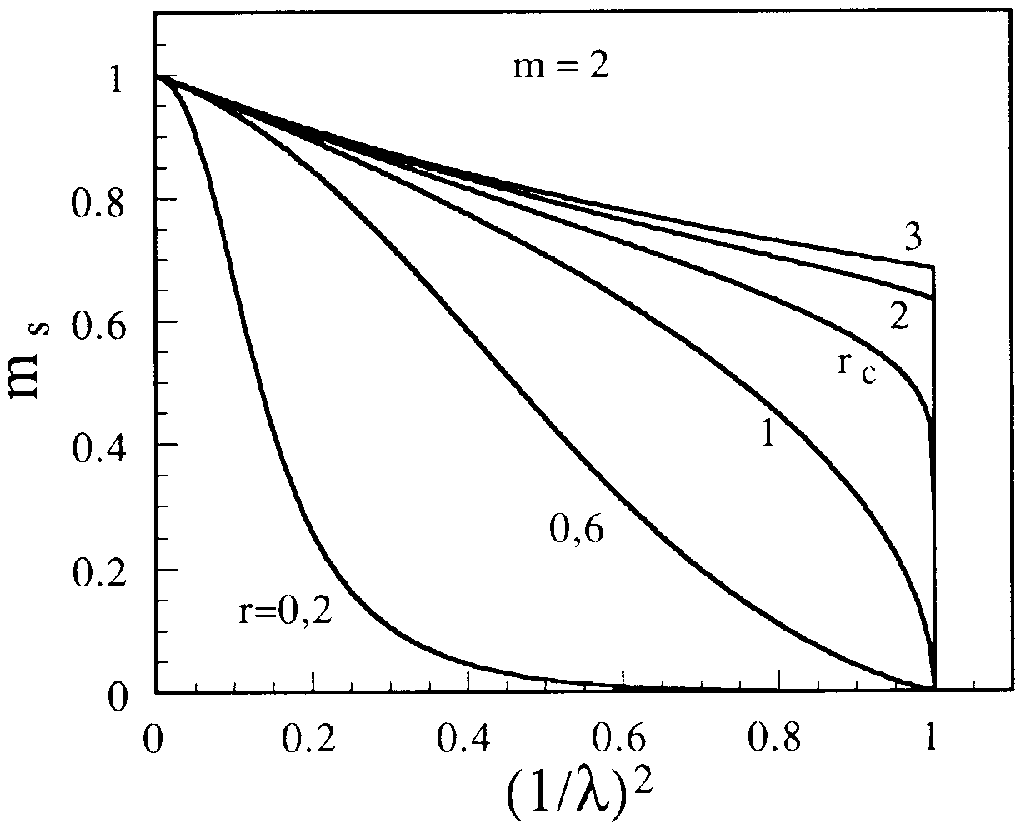}}
\smallskip
\figure{Spontaneous surface magnetization $m_s$ of the Ising quantum
chain with the $m\!=\!2$ Fredholm aperiodic modulation as a function
of the square of the reduced coupling $\lambda_c/\lambda$ for different values
of the coupling ratio $r$. The surface transition is first--order when $r\!>\!
r_c\!=\sqrt{2}$.}
\endinsert
\endgroup
\par}

{\par\begingroup\parindent=0pt\medskip
\epsfxsize=9truecm
\midinsert
\centerline{\epsfbox{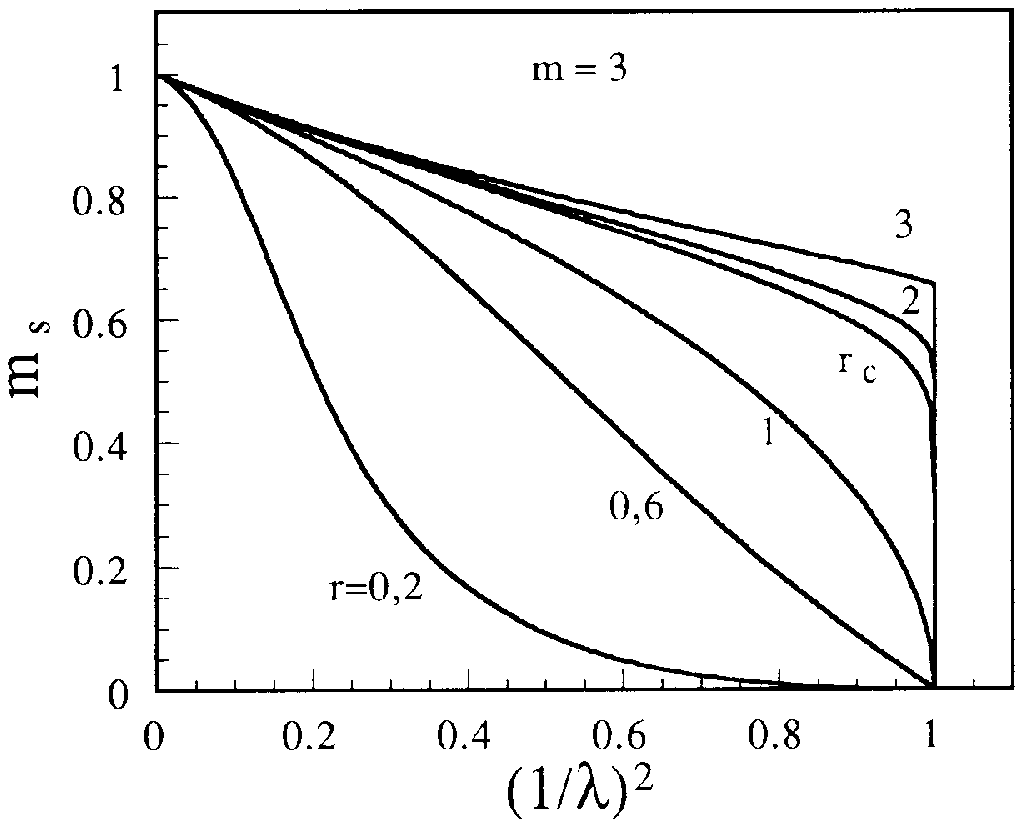}}
\smallskip
\figure{Spontaneous surface magnetization $m_s$ of the Ising quantum
chain with the $m\!=\!3$ Fredholm aperiodic modulation as a function
of the square of the reduced coupling $\lambda_c/\lambda$ for different values
of the coupling ratio $r$. The surface transition is first--order when $r\!>\!
r_c\!=\sqrt{3}$.}
\endinsert
\endgroup
\par}

One easily verifies that the term $l\!=\! n$ of the sum in~(3.10) contains the
powers of $\lambda^{-2}$ from $m^n\!+\!1$ to $m^{n+1}$. Cutting the sum at
$n\!-\!1$, one collects the contribution to the first $L\!=\! m^n$ terms of the
series expansion and, using~(3.7),
$$
\fl S_{L=m^n}(1,r)=2+{m-1\over r^2}\sum_{l=0}^{n-1}\left({m\over r^2}\right)^l=
2+{m-1\over r^2}\left[{1-(mr^{-2})^n\over 1-mr^{-2}}\right].
\eqno(3.11)
$$

Two regimes have then to be considered: when $r\!<\! r_c\!=\!\sqrt{m}$,
$S_{L=m^n}$ behaves as $(mr^{-2})^n$ whereas it is $O(1)$ when $r\!>\! r_c$. It
follows that
$$
\eqalign{
\beta_s&={1\over2}-{\ln r\over\ln m} ,\qquad r\leq r_c=\sqrt{m} ,\cr
\beta_s&=0 ,\qquad\qquad\ \ \ \ \  r>r_c .\cr}
\eqno(3.12)
$$

When $r\!\leq\! r_c$ the surface transition is second order as shown in
figures 1 and 2. The  exponent $\beta_s$ depends on the modulation amplitude,
as
expected with a marginal perturbation. This dependence is shown in figure 3.
The exponent goes to zero linearly at $r_c$ as $(r_c-r)/(2r_c\ln r_c)$.

The vanishing of $\beta_s$ for $r\!>\! r_c$ signals the occurence of
surface order at the bulk critical point. Since the surface is
one--dimensional in the corresponding layered two--dimensional classical
system, it cannot stay ordered when the bulk disorders and the surface
transition is first--order in this regime as shown in figures 1 and 2. The
existence of surface order at $\lambda_c$ and above is linked to the
localization of the eigenvector $\phi_1$. It remains normalizable even at the
critical point when $r\!>\! r_c$.

{\par\begingroup\parindent=0pt\medskip
\epsfxsize=9truecm
\topinsert
\centerline{\epsfbox{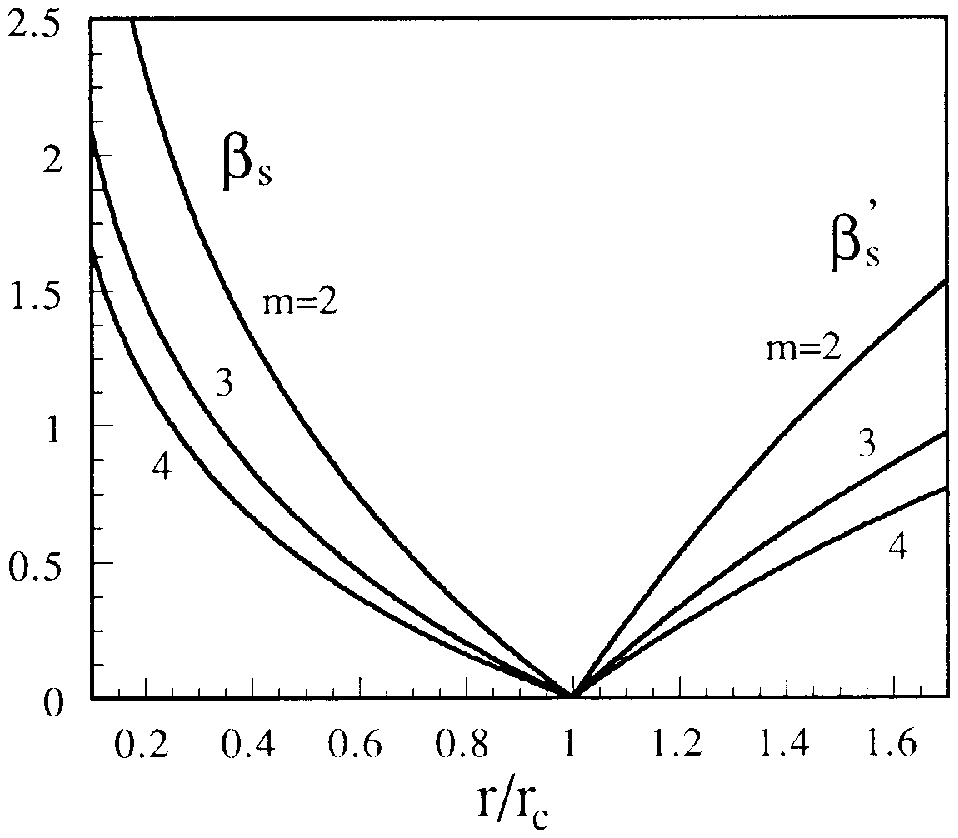}}
\smallskip
\figure{Ising Surface magnetic exponents $\beta_s$ and $\beta'_s$ versus
reduced coupling ratio $r/r_c$ for the generalized Fredholm sequence with
$m\!=\!2$, $3$, $4$. The exponent $\beta'_s$ is associated with the approach
towards the critical surface magnetization when $r>r_c$.}
\endinsert
\endgroup
\par}

The value of the critical surface magnetization $m_{s,c}$ follows
from~(3.5) taking the limit $n\!\to\!\infty$ in~(3.11), which gives:
$$
m_{s,c}=\sqrt{r^2-m\over2r^2-m-1} ,\qquad r\geq r_c .
\eqno(3.13)
$$
The behaviour of $m_{s,c}$ is shown in figure 4. It vanishes with a
square--root
singularity at $r_c$.

{\par\begingroup\parindent=0pt\medskip
\epsfxsize=9truecm
\topinsert
\centerline{\epsfbox{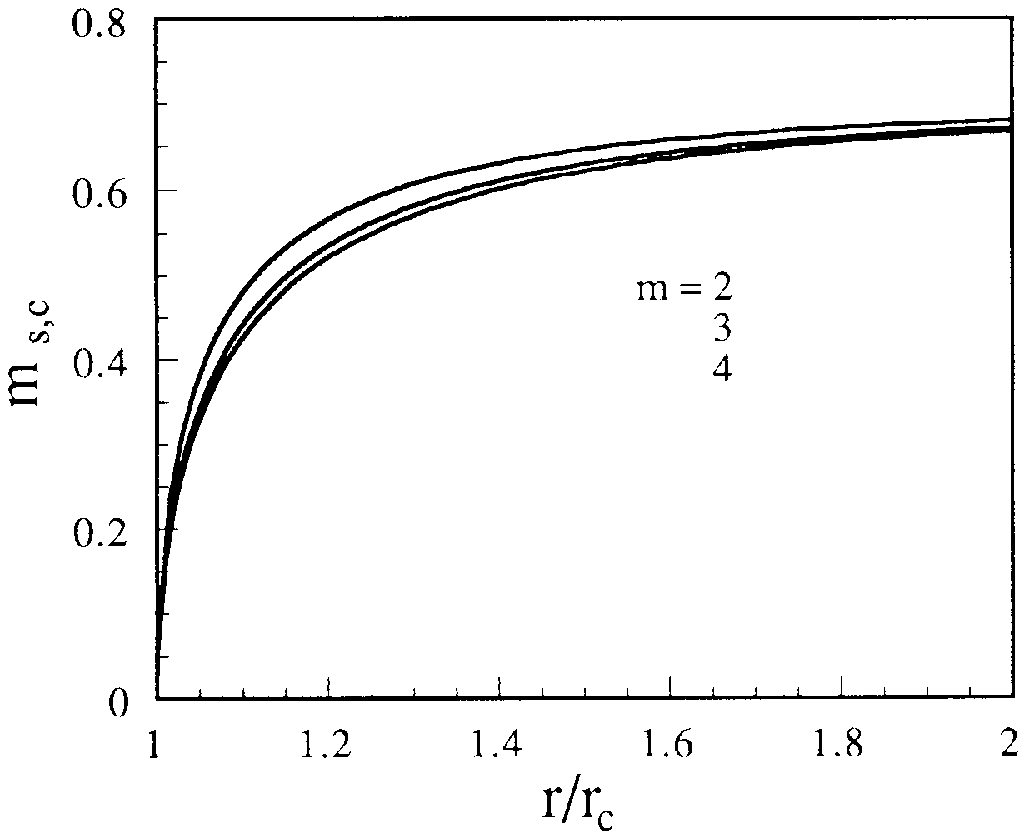}}
\smallskip
\figure{Variation of the critical surface magnetization $m_{s,c}$ with
the reduced coupling ratio $r/r_c$ for the generalized Fredholm sequence with
$m\!=\!2$, $3$, $4$ from top to bottom.}
\endinsert
\endgroup
\par}

The approach to the critical surface magnetization above $r_c$ involves another
surface exponent $\beta'_s$ such that $m_s\!-\! m_{s,c}\!\sim\!t^{\beta'_s}$.
It may be calculated, using the same method as above, by considering the series
expansion $S'_L(1,r)\!=\! S_L(1,r)\!-\! [m_{s,c}]^{-2}$, with $L\!=\! m^n$,
which scales like $L^{-\beta'_s}$. One obtains: $$
\eqalign{
&S'_L(1,r)={m-1\over r^2}\sum_{l=n}^{\infty}\left({m\over r^2}\right)^l
\sim\left({m\over r^2}\right)^n\sim m^{-n\beta'_s} ,\cr
&\beta'_s=2\ {\ln r\over\ln m}-1 .\cr}
\eqno(3.14)
$$
The exponent $\beta'_s$, shown in figure 3, vanishes linearly at $r_c$ as
$(r-r_c)/(r_c\ln r_c)$.

At the critical value of the modulation amplitude, according to~(3.11), the
finite series expansion gives
$$
S_{L=m^n}(1,r_c)=2+{m-1\over m}\ n\sim \ln L ,
\eqno(3.15)
$$
so that the surface magnetization vanishes there with a logarithmic singularity
$$
m_s(r_c)\sim(-\ln t)^{-1/2} .
\eqno(3.16)
$$
The variation of $m_s(r_c)$ with $\lambda^{-2}$ is shown in figures 1 and 2.

\section{Surface energy}
The critical behaviour of the surface energy can be studied
by considering  the finite--size behaviour of the matrix element
$e_s\!=\!\langle\varepsilon\!\mid\!\sigma_1^z\!\mid\!0\rangle$ which does not
contain any regular part and scales like $L^{-x_{e_s}}$ where $x_{e_s}$ is the
dimension of the surface energy density. Here, the state
$\mid\!\varepsilon\rangle$ is the lowest two--fermion eigenstate
$\eta_1^\dagger\eta_2^\dagger\!\mid\!0\rangle$ of the Hamiltonian~(3.1).
Writing $\sigma_1^z$ in terms of diagonal fermions, one obtains $$
e_s=\psi_1(1)\phi_2(1)-\phi_1(1)\psi_2(1)
=(\epsilon_2-\epsilon_1)\phi_1(1)\phi_2(1) \eqno(4.1)
$$
where, in the last expression, we used the relation
$\psi_\nu(1)\!=\!-\epsilon_\nu\ \phi_\nu(1)$ which follows from the second
equation in~(3.2) for the surface components of the eigenvectors.

The matrix element $e_s$ on finite systems with size $L$ of the form $m^n$  was
obtained through  a numerical solution of the eigenvalue problem
$$
\eqalign{
&\lambda_{k-1}\phi_\nu(k-1)+(\lambda_{k-1}^2+1)\phi_\nu(k)+
\lambda_k\phi_\nu(k+1)
=\epsilon_\nu^2\phi_\nu(k) ,\cr
&\phi_\nu(1)+\lambda_1\phi_\nu(2)=\epsilon_\nu^2\phi_\nu(1) ,
\qquad\phi_\nu(L+1)=0 .\cr}
\eqno(4.2)
$$
For small chains $(L\!\leq\!2^{15}$ or $3^9)$, the complete excitation matrix
was diagonalized while for longer chains (up to $L\!=\!2^{17}$ or $3^{10})$,
equation~(4.2) was rewritten as a matrix recursion relation
$$
\left(\matrix{\phi_\nu(k+1)\cr\phi_\nu(k)\cr}\right)=
\left(\matrix{{\epsilon_\nu^2-\lambda_{k-1}^2-1\over\lambda_k}
&-{\lambda_{k-1}\over\lambda_k}\cr
1&0\cr}\right)
\left(\matrix{\phi_\nu(k)\cr\phi_\nu(k-1)\cr}\right)
\eqno(4.3)
$$
and varying $\epsilon_\nu$, we looked for the zeros of $\phi_\nu(L\!+\!1)$,
with $\phi_\nu(0)\!=\!0$ and $\phi_\nu(1)$ arbitrary in the first vector. Once
the zeros corresponding to the two lowest eigenvalues are found, the
corresponding eigenvectors are normalized in order to evaluate~(4.1). The
exponent $x_{e_s}$ is then deduced from the slope at large $L$ in a log--log
plot. The results are shown in figures 5 and 6.

The behaviour of $x_{e_s}$ below $r_c$ can be deduced from finite--size scaling
considerations. Low--energy excitations scale as $L^{-1}$ and due to the
factor $\epsilon_\nu$, the l. h. s. of the first equation in~(3.2) can be
neglected. The leading finite--size behaviours of $\phi_1(1)$ and $\phi_2(1)$
are the same and follow from~(3.4) by cutting the sum at $j\!=\! L$. The
calculation proceeds like for $\beta_s$ and gives
$\phi_1(1)\!\sim\!\phi_2(1)\!\sim\! L^{-x_{m_s}}$ with
$x_{m_s}\!=\!\beta_s\!=\!1/2-\ln r/\ln m$. As a by--product, we recover
$\nu\!=\!\beta_s/x_{m_s}\!=\!1$ for the bulk correlation length exponent, as
expected for a surface perturbation. Collecting these results in~(4.1), we
obtain:
$$
e_s(L)\sim L^{-1-2\beta_s} ,\qquad x_{e_s}=2-2\ {\ln r\over\ln m} ,\qquad r\leq
r_c .
\eqno(4.4)
$$

{\par\begingroup\parindent=0pt\medskip
\epsfxsize=9truecm
\topinsert
\centerline{\epsfbox{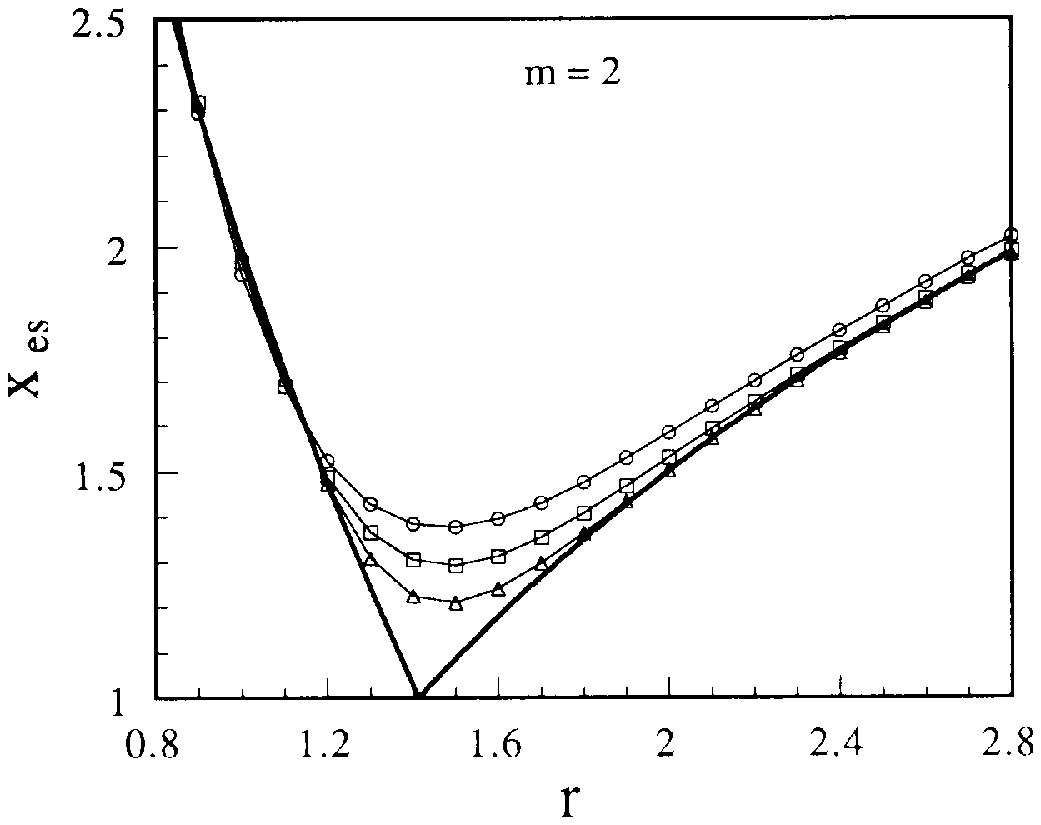}}
\smallskip
\figure{Surface energy density exponent as a function of the coupling
ratio $r$ for the Fredholm sequence with $m\!=\!2$. The points correspond to
finite--size scaling results on chains with sizes of the form $2^{n}$ up to
$L\!=\!2^{10}$, $2^{12}$, $2^{17}$ from top to bottom. The heavy line gives
the conjectured analytical result.}
\endinsert
\endgroup
\par}

{\par\begingroup\parindent=0pt\medskip
\epsfxsize=9truecm
\midinsert
\centerline{\epsfbox{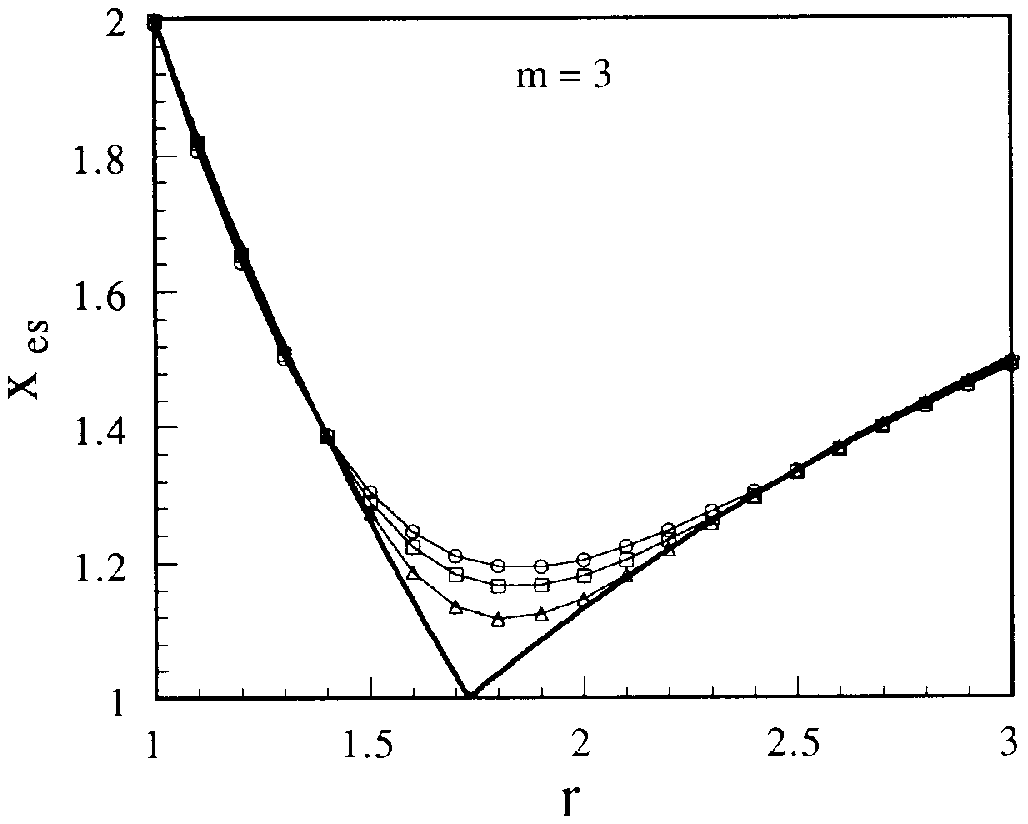}}
\smallskip
\figure{Surface energy density exponent as a function of the coupling
ratio $r$ for the generalized Fredholm sequence with $m\!=\!3$. The points
correspond to finite--size scaling results on chains with sizes of the form
$3^{n}$ up to $L\!=\!3^{7}$, $3^{8}$, $3^{9}$ from top to bottom. The heavy
line gives the conjectured analytical result.}
\endinsert
\endgroup
\par}

When $r\!>\! r_c$ one finds different size--dependences for the first and
second excitations as well as for the corresponding components of the
eigenvector. Due to surface ordering, the first excitation vanishes anomalously
as $L^{-2\ln r/\ln m}$, i. e. quicker than the higher ones with the usual
$L^{-1}$ behaviour. Equation~(3.3) can still be used to calculate the
size-dependence of $\phi_1(1)$ which, following the same steps as in the last
section, gives  $m_s(L)\!={\rm constant\ term}+ L^{-x'_{m_s}}$ where
$x'_{m_s}\!=\!\beta'_s$. The term on the~l.~h.~s. in~(3.2) can no longer be
neglected for $\phi_2(1)$. The size--dependence $L^{1/2-\ln r/\ln m}$ was
obtained numerically. Together, this gives:
$$
e_s(L)\sim L^{-1-x'_{m_s}/2} ,\qquad x_{e_s}={1\over2}+{\ln r\over\ln m}
,\qquad r>r_c .
\eqno(4.5)
$$

One may notice that the anomalous behaviour of the first excitation is
in agreement with a scaling theory for first--order line
transitions involving an irrelevant variable and leading to a
$L^{1-2x_{e_s}}$ size--dependence (Igl\'oi and Turban 1993).
Numerical results for the surface exponent $x_{e_s}$ in figures 5 and 6,
converge smoothly with increasing size towards the conjectured analytic
expressions.

\section{Discussion}
As already mentioned in the introduction, the aperiodic surface extended
perturbation treated here, displays some similarities with the Hilhorst--van
Leeuwen model (Hilhorst and van Leeuwen 1981) in which the coupling, at a
distance $k$ from a free surface, takes the form
$$
\lambda_k=\lambda\left(1+{\alpha\over k^{y}}\right).
\eqno(5.1)
$$
For the Ising model in $1+1$ dimensions, the marginal case corresponds to a
decay exponent $y\!=\!1$. The critical surface order appears at
$\alpha_c=1/2$ and the surface critical exponents have the following
dependence on the perturbation amplitude:
$$
\eqalign{
\beta_s&=x_{m_s}={1\over2}-\alpha ,\qquad
x_{e_s}=2-2\alpha ,\qquad\alpha\leq\alpha_c ,\cr
\beta'_s&=x'_{m_s}=2\alpha-1 ,\qquad x_{e_s}
={1\over2}+\alpha ,\qquad\alpha>\alpha_c .\cr}
\eqno(5.2)
$$
A comparison with previous results in equations~(3.12), (3.14), (4.4)
and~(4.5) shows that the exponents of the aperiodic system are recovered
through
the following correspondence:
$$
\alpha\to{\ln r\over\ln m} .
\eqno(5.3)
$$
In the Hilhorst--van Leeuwen model, the integrated relative perturbation at a
distance $L$ is given by:
$$
\sum_{k=1}^L{\lambda_k-\lambda\over\lambda}=\alpha\sum_{k=1}^L{1\over k}
\simeq\alpha\ln L .
\eqno(5.4)
$$
On the average, the same logarithmic dependence is found with the
aperiodic perturbation for which, on a chain with length $L\!=\! m^n$ according
to~(2.8), the corresponding quantity is
$$
\sum_{k=1}^L{\lambda_k-\lambda\over\lambda}=(r-1)\ n_L={r-1\over\ln m}\ \ln L .
\eqno(5.5)
$$
This would suggest the identification $\alpha\!=\!(r-1)/\ln
m$ which, actually, is only valid up to the first order in $r\!-\!1$, i. e.
for a weak modulation. The fluctuations around the average logarithmic
behaviour modify this expression for a stronger perturbation. The correct
identification can be obtained by considering the finite--size behaviour of
the critical coupling which follows from equation~(3.6) with
$$
\ln\left[\lambda_c(L)\right]=-\ln r\ {n_L\over L}=-{\ln r\over\ln m}\ {\ln
L\over L}
\eqno(5.6)
$$
for the aperiodic system, whereas
$$
\ln\left[\lambda_c(L)\right]=-{1\over L}\sum_{k=1}^L\ln\left(1+{\alpha\over
k}\right)\simeq-\alpha\ {\ln L\over L}
\eqno(5.7)
$$
in the Hilhorst--van Leeuwen model.

To conclude, one may notice that an experimental realization of the
Hilhorst--van Leeuwen model with its $1/k^y$ decay of the couplings appears
difficult while its aperiodic counterpart, studied here in the marginal case,
could be more easily obtained (at least in three dimensions) using appropriate
sequences in multi--layer systems.

\ack
LT is indebted to Ferenc Igl\'oi for stimulating discussions.
DK and GP thank Bertrand Berche for help in the numerical work which
was supported by CNIMAT under project No 155C93b.

\references

\refjl{Au--Yang H and McCoy B M 1974}{\PR\ {\rm B}}{10}{886}

\refjl{Benza G V 1989}{Europhys. Lett.}{8}{321}

\refjl{Ceccato H A 1989a}{\PRL}{62}{203}

\refjl{Ceccato H A 1989b}{\ZP\ {\rm B}}{75}{253}

\refjl{Dekking M, Mend\`es--France M and van der
Poorten A 1983}{Math. Intelligencer}{4}{130}

\refjl{Doria M M and Satija I I 1988}{\PRL}{60}{444}

\refbk{Dumont J M 1990}{Number Theory and Physics, Springer Proc. Phys.}
{vol 47 ed. J M Luck, P Moussa and  M Waldschmidt (Berlin: Springer) p 185}

\refjl{Garg A and Levine D 1987}{\PRL}{59}{1683}

\refjl{Godr\`eche C, Luck J M and Orland H J 1986}{J. Stat. Phys.}{45}{777}

\refjl{Grimm U and Baake M 1994}{J. Stat. Phys.}{74}{1233}

\refjl{Harris AB 1974}{\JPC}{7}{1671}

\refjl{Henkel M and Patk\'os A 1992}{\JPA}{25}{5223}

\refjl{Henley C L and Lipowsky R 1987}{\PRL}{59}{1679}

\refjl{Hilhorst H J and van Leeuwen J M J 1981}{\PRL}{47}{1188}

\refjl{Igl\'oi F 1986}{\JPA}{19}{3077}

\refjl{\dash 1988}{\JPA}{21}{L911}

\refjl{\dash 1993}{\JPA}{26}{L703}

\refjl{Igl\'oi F, Peschel I and Turban L 1993}{Adv. Phys.}{42}{683}

\refjl{Igl\'oi F and Turban L 1993}{\PR\ {\rm B}}{47}{3404}

\refjl{\dash 1994}{Europhys. Lett.}{27}{91}

\refjl{Jordan P and Wigner E 1928}{\ZP}{47}{631}

\refjl{Kogut J 1979}{\RMP}{51}{659}

\refjl{Langie G and Igl\'oi F 1992}{\JPA}{25}{L487}

\refjl{Lieb E H, Schultz T D and Mattis D C 1961}{\APNY}{16}{406}

\refjl{Lin Z and Tao R 1990}{\PL}{150A}{11}

\refjl{\dash 1992a}{\JPA}{25}{2483}

\refjl{\dash 1992b}{\PR\ {\rm B}}{46}{10808}

\refjl{Luck J M 1993a}{J. Stat. Phys.}{72}{417}

\refjl{\dash 1993b}{Europhys. Lett.}{24}{359}

\refjl{McCoy  B M and Wu T T 1968}{\PR}{176}{631}

\refjl{Okabe Y and Niizeki K 1988}{\JPSJ}{57}{1536}

\refjl{\dash 1990}{\JPA}{23}{L733}

\refjl{Peschel I 1984}{\PR\ {\rm B}}{30}{6783}

\refjl{Pfeuty P 1979}{\PL}{72A}{245}

\refbk{Queff\'elec M 1987}{Substitution Dynamical
Systems--Spectral Analysis, Lecture Notes in Mathematics}{vol 1294
ed. A Dold and B Eckmann (Berlin: Springer) p 97}

\refjl{Sakamoto S, Yonezawa F, Aoki K, Nos\'e S and
Hori M  1989}{\JPA}{22}{L705}

\refjl{Schultz T D, Mattis D C and Lieb E H 1964}{\RMP}{36}{856}

\refjl{S\o rensen E S, Jari\'c M V and Ronchetti M 1991}{\PR\ {\rm
B}}{44}{9271}

\refjl{Tracy C A 1988}{\JPA}{21}{L603}

\refjl{Turban L and Berche B 1993}{\ZP\ {\rm B}}{92}{307}

\refjl{Turban L, Igl\'oi F and Berche B 1994}{\PR\ {\rm B}}{49}{12695}

\refjl{Turban L, Berche P E and Berche B 1994}{\JPA}{27}{6349}

\refjl{You J Q, Yan J R and Zhong J X 1992}{\JMP}{33}{3901}

\refjl{Zhang C and De'Bell K 1993}{\PR\ {\rm B}}{47}{8558}

\vfill\eject
\bye